\documentclass[aps,prd,superscriptaddress,showpacs,preprintnumbers]
{revtex4}

\usepackage{amssymb, graphics, setspace}
\usepackage{pgfplots}
\usepackage{textcomp}
\usepackage[caption=false]{subfig}

\usepackage{amsmath}
\usepackage{commath}

\usepackage{graphicx,bm}
\usepackage{url}
\usepackage{float}
\usepackage{textcomp}
\usepackage{verbatim}
\begin{document}

\newcommand{\dlt}{\bigtriangleup}
\newcommand{\beq}{\begin{equation}}
\newcommand{\eeq}[1]{\label{#1} \end{equation}}
\newcommand{\insertplot}[1]{\centerline{\psfig{figure={#1},width=14.5cm}}}

\parskip=0.3cm


\title{Worldlines and social networks}


\author{L\'aszl\'o Jenkovszky}
\affiliation{Bogolyubov Institute for Theoretical Physics (BITP),
 National Academy of Sciences of Ukraine \\14-b, Metrologicheskaya str.,
Kiev, 03143, UKRAINE; jenk@bitp.kiev.ua}

\begin{abstract}
Familiar laws of physics are applied to study human relations, modelled by their world lines (worldlines, WLs) combined with social networks. 
We focus upon the simplest, basic element of any society: a married  couple, stable due to the dynamic balance between attraction and repulsion. By building worldlines/worldsheets, we arrive at a two-level coordinate systems: one describing the behaviour of a string-like binary system (here, a married couple), the other one, external, corresponding to the motion of this couple in the medium, in which the worldline is embedded, sweeping there a string-like sheet or brane.  
  The approach is illustrated by simple examples (semi-quantitative toy models) of worldlines/sheets, open to further extension, perfections and generalization. World lines (WLs) are combined with social networks (SN). Our innovation is in the application of basic physical laws, attraction and repulsion to human behaviour. 
Simple illustrative examples with empirical inputs taken from intuition and/or observation are appended. This is an initial attempt, open to unlimited applications.     

\end{abstract}

\maketitle


\section{Introduction} \label{Sec:1}
The behaviour of individuals or groups of people analyzed by means of mathematical physics, appended by empirical data (observations) is attracting researchers working in various fields of natural sciences:  mathematics, biology, sociology, economy etc., see e.g. \cite{Buchanan, Barabasi, Lovasz, Galam, Perc, Perc1}. A branch related to sociology is called social physics or sociophysics \cite{Sociophysics, Sociology}. It investigates various phenomena by analogy with astronomical, physical, chemical, and physiological phenomena. Social physics refers also to using big data analysis and the mathematical laws to understand collective effects of human crowds. The  basic idea is that data about human activity contain mathematical patterns that are characteristic of social interactions as well. 

The application of methods of mathematical physics in chemistry, biology, physiology, sociology and related area is not new. In the present paper I try to advance further by including innermost motivations of human behaviour and/or groups of persons. 

I adhere to the philosophical principal of dualism by which life is based on balance between two competing components, "good" (love, interest, sympathy, empathy etc.) and "bad" (disgust, hate, antipathy, evil etc.). Alternatives to dualism are monism, by which only a singly category exists, or pluralism, implying a multitude of categories in nature. We chose dualism (Sec. \ref{Sec:Conjugal}) for its universality, simplicity and affinity to the basic laws of physics.    

The idea that mental and bodily events are coordinated, without causal interaction between them is known as {\it philosophical parallelism}. It assumes correlation of mental and bodily events, but denies any direct causal relation between mind and body. Accordingly, mental and bodily phenomena are independent, yet inseparable. 
  Psychophysical parallelism \cite{Chisholm} is a third possible alternative regarding the relation between mind and body, between interaction (e.g., dualism) and one-sided action (e.g., materialism, epiphenomenalism). It is a theory related to dualism suggesting that although there is correlation between mental and physical events there is no causal connection. The definition of and relation between {\it body and soul} is a delicate issue, not easy to fix rigorously or unambiguously. 


In any case, I believe that part of the universe, namely its materialistic component may be conceived or, at least approached by means of mathematical physics, while the other one, spiritual belongs to theology. I do not enter that area, instead try to approach the interface region between the two by using general and flexible physical laws and models of interaction. These models are based on attraction and repulsion with free parameters adjusted to empirical data, intuition and guesses suggested by great artworks.  By the latter I mean the treasury of world art, accumulated in works of philosophers, artists, writers, poets and musicians. Great artists were able to unveil the past and intuitively reveal the future.

A typical problem in specifying the border and transition region between matter and spirit is in understanding emotions (feelings). Physiologists attempted to find the origin and nature of feeling in experiments with animals. Popular are  discussions connected with the pleasure and searches of relevant centres in animals and human beings. Brain stimulation reward (BSR), discovered by James  Olds and Peter Milner \cite{Olds} is a pleasurable phenomenon generated by stimulation of specific brain regions. Profound observations of anlimals' internal world ({\it ethology}) and mysterious telepath phenomena can be found in Refs. \cite{Lorenz1, Lorenz2}.


Human souls are battlefields between body (physiology) and spirit (divine), subject of literature masterpieces such as L. Tolstoy's "Father Sergius" \cite{Tolstoy}. I will not penetrate the "other side", instead try to model the behaviour of human beings based on observations, literature, intuition and common sense. I try to demonstrate that human relations may follow simple laws of attraction and repulsion appended by observations (empiric). 

At this point it is appropriate to cite Emmanuel Kant \cite{Kant}: "Two things fill the mind with ever new and increasing admiration and awe, the more often and steadily we reflect upon them: the starry heavens above me and the moral law within me. I do not seek or conjecture either of them as if they were veiled obscurities or extravagances beyond the horizon of my vision; I see them before me and connect them immediately with the consciousness of my existence." 

The aim of the present paper is animation of  world lines and social networks. By "animation" I mean inspiration, i.e. endowing familiar mechanical, statistical or mathematical constructions (world lines and social networks) with footprints of spirit, absent from machines but present in human behaviour. 


The paper is organized as follows: in Sec. \ref{Sec:Wlines} I introduce World lines, to be combined with networks in Sec. \ref{Sec:WL_Net}. In Sec. \ref{Sec:Binary} I overview various binary systems and their interrelation, with Sec. \ref{Sec:Interrelate}, preparing the ground for the central part of the paper, Sec. \ref{Sec:Conjugal} modelling the dynamics of particular binary systems, namely those of married couples.    


\section{Worldlines} \label{Sec:Wlines} 
Worldlines (WL) \cite{WL1}, called also time-space geography are three-dimensional plots in which two dimensions are spatial (we use flat, Euclidean space), the third coordinate being time. 

The existence of the fourth, time dimension, apart from those spacial was intuited, well before Lorentz, Poincar\'e and Minkovski by great Greeks, followed by Spinoza, Kant, I. Newton and, last but not least, Herbert Wells in his {\it Time Mashine}. Less familiar, moreover unkonwn, are the related ideas of the Russian philosopher Aksionov, available in Refs. \cite{Aksionov, UFN}.

When Newton formulated his theory of gravity, he assumed time to be linear, with unchangeable rate of flow. He assumed space to the absolute, unchanging, and Euclidean: ’the divine sensorium’ \cite{Barrow}.

\begin{figure}[H] 	
	\centering	
	\includegraphics[scale=0.7]{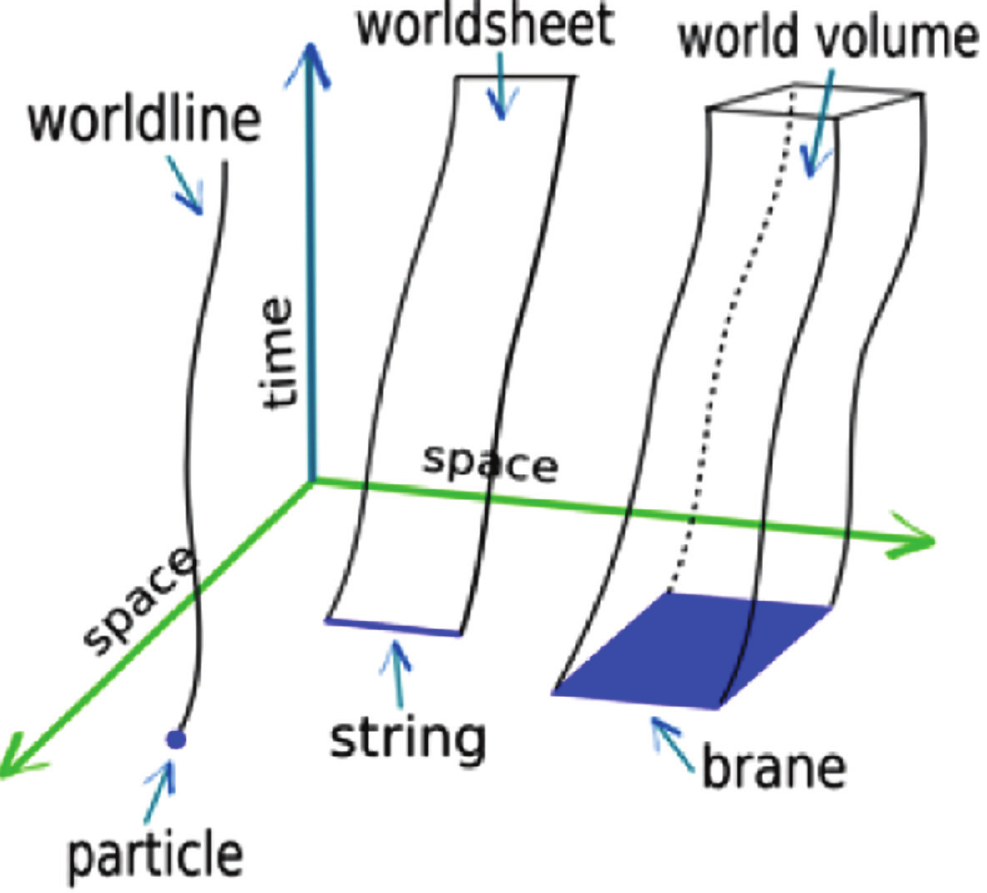}
		\caption{World lines and branes: from the WL of a point-like object (leftmmost), via a two-dimensional sheet (propagating rubber band, {\it i.e.} a string) to a multi-dimensional brane.}   
	\label{Fig:WL1}
	\end{figure}

	\begin{figure}[H] 	
	\centering
	\includegraphics[scale=1.0]{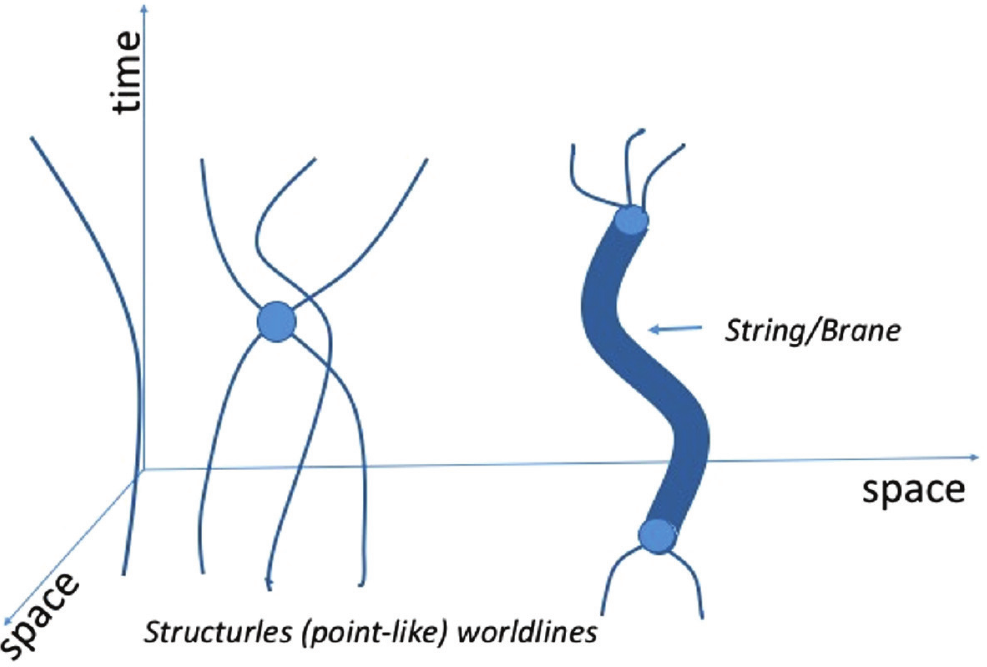}
	\caption{Worldlines (WL). Leftmost is a dimensionless line; crossing lines (next to the right) may or may not interact. The rightmost object, emerging from the merge of two structureless lines has finite dimensions; it is called strip (rubber band), tube or brane.}   
	\label{Fig:WL}
	\end{figure}

	Time geography or time-space geography is an interdisciplinary merge of spatial and temporal events. Time geography is a framework and visual language in which space and time are basic dimensions to analyse dynamic processes. Time geography was originally developed by geographers, but now it is used in various fields including anthropology, environmental science etc. Since the 1980s, time geography is used also by researchers in biological and social sciences, and in interdisciplinary fields.

Benjamin Bach with his colleagues \cite{Bach} have generalized the space-time cube towards a framework for temporal data visualization applicable to all data that can be represented in two spatial dimensions plus time.

Worldlines are perfect tools to illustrate biographies \cite{Gamow, Jenk}. Constructing and drawing worldlines of known persons, based on documents and/or (auto)biographies, combined with their genealogical lines is an amusing and useful exercise.  

\begin{figure}[H] 
	\centering
	\includegraphics[scale=0.8]{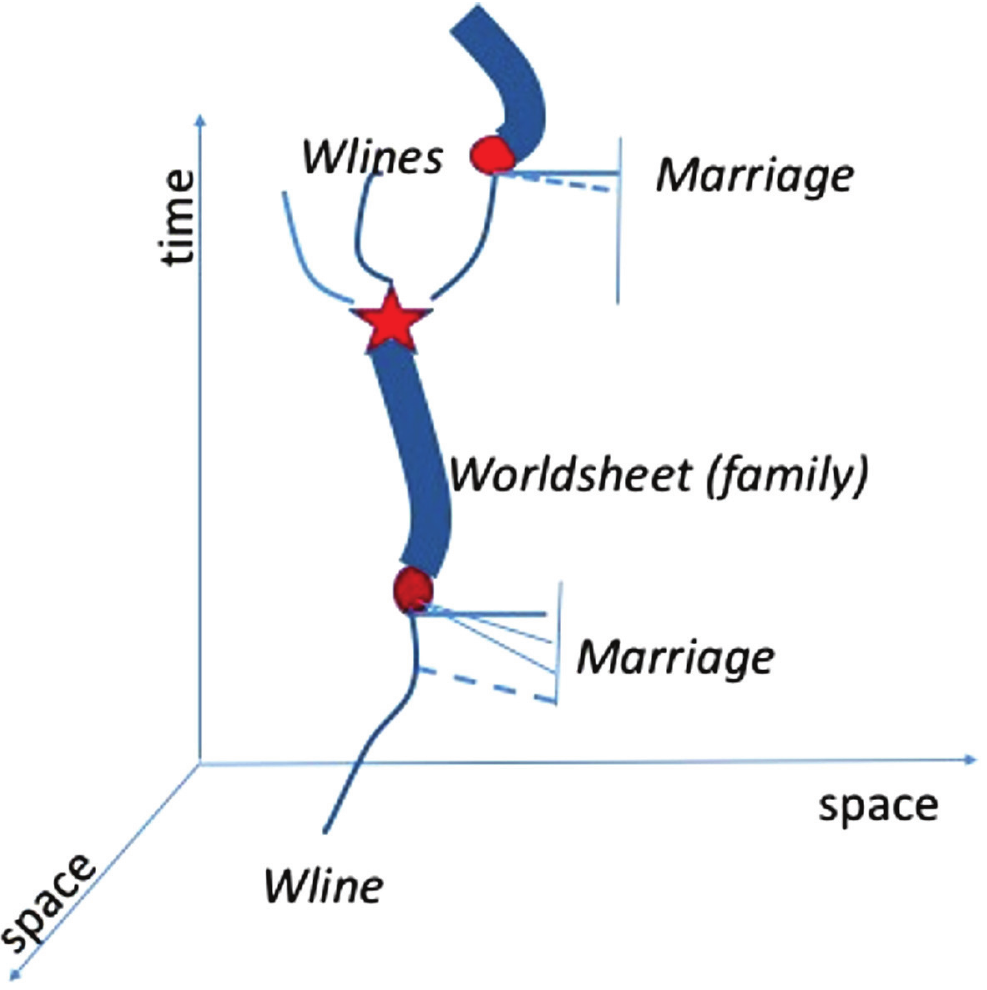}
	\caption{Worldlines appended by elements of a "marriage network" (right margin, to be expanded in Sec. \ref{Sec:Hall}). Such worldlines may evolve also in genealogical graphs, alluded to at the bottom and top of the present figure.}
	\label{Fig:Marriage}
\end{figure}

\begin{figure}[H] 
	\centering
	\includegraphics[scale=1.4]{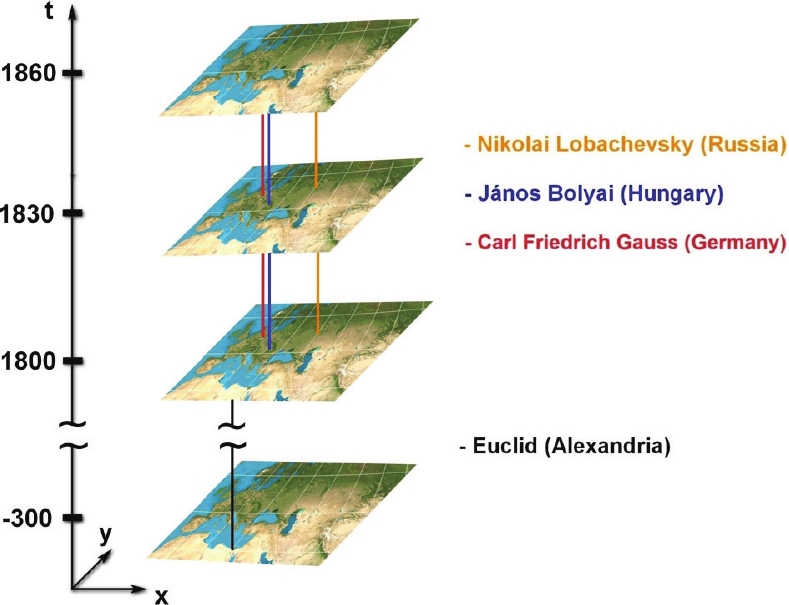}
	\caption{Straightened WLs of Euclid, J\'anos Bolyai, C.F. Gau{\ss} and N.I. Lobachevsky. Three genii who discovered the new, non-Euclidean world,
living on the same continent, nearly at the same time, never met \cite{Jenk}.}
	\label{Fig:BGL}
\end{figure}

A historical example with simplified (straightened) worldines (WL) of four actors (EBGL) is shown in Fig. \ref{Fig:BGL}, Ref. \cite{Jenk}. 

Worldlines (time geography) may be useful to visualise events or, extended/appended by social networks, in analysing data (e.g. in history) and in making predictions (e.g. in sociology) by extrapolation. 

WLs themselves offer an infinity of options and applications, e.g. by replacing lines with extended objects - strips, sheets, bands, tubes, branes. (NB: A brane is a physical object that generalizes the notion of a point particle to higher dimensions.).   2) increasing the number of lines/tubes, including continuum (merge of WLs). With the advent of computational and storage capacities, infinitely large manifolds of WLs, coming close and interacting multiply, evolving towards a continuous, three-dimensional bulk of world history, parametrized by relevant computer codes may be realized in the near future.    

A particular extension of world lines are the phase- and configuration spaces. In classical mechanics, {\it phase space} is the space of all possible states of a system. Remind that the state of a mechanical system is determined by the position $p$ and momenta $q$ of its constituents, where $p$ and $q$ predict the further evolution of the system at any time, provided the laws governing the motion of these objects are known. 

In configuration space, the parameters that define the configuration of the system are called generalized coordinates and the vector space defined by these coordinates is called configuration space. For example, the position of a single particle moving in ordinary Euclidean three-dimensional space is defined as $q=(x,y,z)$ and its configuration space is $R^3$. A particle may be constrained to move within a specified manifold. For example, if the particle is constrained by a rigid bond, free to swing about the origin, it is constrained to lie on a sphere. For $n$ disconnected, non-interacting point-like particles, the configuration space is $R^{3n}$. A word of warning: in life sciences/sociophysics, mechanical momentum should be replaced by a relevant variable.

In this sense, the predictive power of the phase- or configuration spaces is the same as that of a world line, but it is a convenient and powerful technical tool, especially in the case of a large number of objects. The formalism of the phase- and configuration spaces offers huge perspectives in social science studies, provided we know the laws governing human beings or societies. {\it A priory} we do not. In the present paper I try to guess and model these laws and regularities, to be verified empirically. 

Worldlines and social networks are different. While WLs evolve in time along certain trajectories, as shown {\it e.g.} in Figs. \ref{Fig:WL1}--\ref{Fig:BGL}, networks are static. Time dependence, and more details, such as the "price" of a vertex {\it etc.}, may be introduced in networks. If so, the time dependence becomes hierarchic: "internal" within the network and "external" along the WL, as in Sec. \ref{Sec:Conjugal} and Fig. \ref{Fig:Embed}, obeying statistics and topology. Still, they have some common features. With genealogy trees included, WLs acquires many features of a social network (mind the arrow of time!), hence one may look for a particular {\it duality} by interchanging time (vertical orientation) and the (horizontal) spatial coordinate ("interaction range"), remembering however of the uniqueness of the time arrow.     

In perspective, worldlines may play an important role in descriptive history. By this I mean a detailed panoramic view of the evolution of mankind including WLs of individuals and groups, societies etc., as well as their interaction/intersection at various levels and forms. The realization of such a huge "bank of world lines" technically was incredible in the past, but now, with the advent of huge computation and storage capacities, it may be realized! 


In the present paper I consider WLs of a single person or a couple -- building block of our societies. This will be useful when generalized to more complex systems, their interaction, collective effects etc., all that now realizable.

\section{Worldlines and networks, arrow of time} \label{Sec:WL_Net}

Networks \cite{Neural, Konig, Net} are studied and used in mathematics, computer science, geography and other fields of science. Random networks were proposed by Erd\H{o}s and R\'enyi \cite{Renyi} at the end of 1950s. The interest was renewed and reinforced after the discovery  by Albert and Barab\'asi \cite{Barabasi} of strong heterogeneities. 

\begin{figure}[H] 
	\label{fig:Pol}
	\centering
	\includegraphics[scale=1.0]{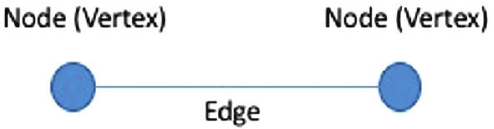}
	\caption{Simplest, primitive net: a binary system (see Sec.\ref{Sec:Conjugal}).}	
\end{figure}  

Network science studies also complex networks such as telecommunication networks, computer networks, biological networks, cognitive and semantic networks, and social networks, considering distinct elements or actors represented by nodes (vertices) and the connections between the elements as links (edges), see {\it e.g.} \cite{Net} and references therein. The field exploits graph theory from mathematics, statistical mechanics from physics, data mining and information visualization from computer science,  and social structure from sociology. D\'enes K\"onig \cite{Konig} was among the pioneers. Probabilistic theory in network science was developed in Paul 
Erd\H{o}s and Alfr\'ed R\'enyi's papers on random graphs \cite{Renyi}.

Albert-L\'aszl\'o Barab\'asi and R\'eka Albert \cite{Barabasi} developed the scale-free network which is a loosely defined network topology that contains hub vertices with many connections, that grow in a way to maintain a constant ratio in the number of the connections versus all other nodes.  Network models serve as a tool in understanding interactions within empirical complex networks.

The Erd\H{o}s –- R\'enyi model \cite{Renyi} is used for generating random graphs in which edges are set between nodes with equal probabilities. It can be used in the probabilistic method to prove the existence of graphs satisfying various properties. The Barab\'asi –- Albert model (BA) \cite{Barabasi} is a random network model used to demonstrate a preferential attachment or a "rich-get-richer" effect (see also \cite{Buchanan}).


A successful social network model is that of Galams \cite{Galam}. Their work focuses on the dynamics of group decision making and how minority opinions can influence public opinion.

Bipolarity, ({\it e.g.} Western "democracy" vs. Eastern "administrative command system") has a parallel with the title of the present paper: while {\it networks} correspond to democracy, {\it worldlines} are hierarchic. Funding of science is an example: centralized, vertical hierarchic funding, typical of the ex-Soviet Union is opposite to the horizontal system of grants, based on unbiased pear referees system (network!), provided it is free of corruption and conflicts (coincidence, correlation) of interests.       

Networks are widely used in estimating citation indices, that became important in scientometrix in deciding about the financial support of a researcher, group or institution. A citation network is a kind of social network that can be represented as a direct graph with nodes representing papers {P1, …, Pn} and edges e(Pi, Pj) between two nodes Pi and Pj denoting a co-citation relationship \cite{Perc1}, when the paper Pi cites paper Pj. The number of citations of scientific articles is becoming one of the most important measures of scientific impact and quality. Hence, the authors are trying to obtain as many citations as possible for their works by creating corrupted citation cartels, where members cite each other in order to increase their own number of citations.

 Besides the structure of interactions within the networks, of interest is also the interaction between networks. This aspect was studied in Ref. \cite{Perc} and papers quoted therein. 


As repeatedly emphasized, we combine ascending (arrow of time!) WLs with horizontal networks. Symmetries with respect to space, $P$, time, $T$ and charge conjugation, $C$ and their combinations play an important role in the microworld. In a fantastic scenario, such symmetries may be used in  sociophysics as well. A simple example of such synthesis is shown in Fig. \ref{Fig:Marriage}, where the horizontal lines (marriage) on the right margin are hinted. They will be discussed in more details in Sec. \ref{Sec:Hall}, referring to Hall's marriage theorem.    

{\it  Music} is a symbiosis of horizontal networks and vertical world lines (evolution with time): while harmony (key, chord, orchestration) is horizontal, melody and rhythm correspond to vertical evolution along the time arrow. The merge of the two produces {\it symphony}. Plastic art, painting, architecture, photo art are "frozen  music".



 \section{Binary systems (dyads, dipoles, biparticle graphs)} \label{Sec:Binary}
In this Section I specify the notion of binary systems. They may imply individuals, families, companies, countries, nations. Binary systems form the basis of further generalizations to "many-body" systems, big numbers, collective phenomena, etc.

With dynamical equilibrium in mind, I rely on models of binary interactions known in physics. In the microworld (e.g. in the "standard theory" of basic, {\it i.e.} electro-week and strong interactions), stable are systems formed by two elementary constituents of opposite charge, as in the hydrogen atom, made of an electron and a proton or in a meson, made of a quark and antiquark. Quarks are bound by strings forming  dipoles in mesons or  triangles (or "Mercedes" stars) in baryons. Different is gravitation where all massive objects attract, no "antigravity"! 
 
 The emergence and evolution of coupling in a binary system is an essential ingredient in our analysis. It depends on many factors: internal motivation, external influence and many more. Three types of motivations may induce two-body correlations: 

\begin{itemize}
\item{\bf Confounding}. A motivation for correlation between actions of adjacent agents in a social network is external influence from elements in the environment. Mathematically, this means that there is a confounding variable $X$, and both the network $G$ and the set of active individuals $W$ come from distributions correlated with $X$. This is in contrast with the influence model, defined below.

\item{\bf Influence}. An obvious explanation for social correlation is social influence. Mathematically, this can be modelled as follows: first, the graph $G$ is drawn according to some distribution. Then, in each of the time steps $1, . . ., t$, each non-active agent decides whether to become active. The probability of becoming active for each agent $u$ is a function $p(x)$ of the number $x$ of other agents $v$ that have an edge to $u$ and are already active. 

\item{\bf Homophile}. The third and the most obvious tendency of individuals to choose partners is based on similarity of characteristics (homophile, see Sec. \ref{Sec:AppI}). It leads to correlation between the actions of adjacent nodes in a social network. Mathematically, the set $W$ of active nodes is first selected according to some distribution, and then the graph $G$ is picked from a distribution that depends on $W$.
\end{itemize}


\subsection{Measuring social correlations} \label{Sec:Interrelate}

Social correlation is a well-known phenomenon. Formally, this means that for two nodes $u$ and $v$ that are adjacent in $G$, the events that $u$ becomes active is correlated with $v$ becoming active. There are three primary explanations for this phenomenon: homophile, the environment (or confounding factors), and social influence. 
A logistic function with the logarithm of the number of friends as the explanatory variable provides a good fit for the probability. Therefore, one uses the logistic function with this variable, that is one estimates the probability $p(a)$ of activation for an agent with a already-active partner as follows:
\begin{equation}\label{1}
p(a) = \frac{e^{\alpha \ln(a+1)+\beta}}{1 + e^{\alpha \ln(a+1)+\beta}},
\end{equation}
where $\alpha$ and $\beta$ are parameters. 
The parameter $\alpha$ measures social correlation: a large value of $\alpha$ indicates a large degree of correlation.

\section{Supply and demand; optimal pairing} \label{Sec:Hall}

Optimization of marriages is a popular exercises in network theory, especially that "marriage" may imply variants. Huge literature exists on the subject, see e.g. \cite{Marriage, Matching, Hall, Dilworth}  and earlier references therein.

A popular model is based on the so-called Hall's marriage theorem \cite{Hall} resulting from the combinatory that specifies distinct elements to be chosen from overlapping finite sets of elements. It is equivalent to several theorems in combinatory, including that of Dilworth \cite{Dilworth}. The name comes from an application to matchmaking, given a list of potential matches among an equal number of brides and grooms. The theorem gives a necessary and sufficient condition on the list for everyone to be married to an optimal match. Also, this is an example of the efficiency of the social networks.
 
\subsection{Application to marriage, business partners, etc.}
Suppose there is a certain number of women and men wanting to get married to someone of the opposite sex. Suppose that the women each have a list of the men they would like to marry, and that every man would like to marry any woman who is happy to marry him, and that each person can only have one spouse.

The best known mathematical result is Hall's marriage theorem \cite{Hall}. It says that men and womenfolk can all be paired iff the following marriage condition holds: if in any group of women, the total number of men who are acceptable to at least one of the women in the group is greater than or equal to the size of the group.

It is clear that this condition is a necessary one. Hall's marriage theorem says that it is also sufficient.
 
 This is a wide and popular subject, related to different areas of life, not only marriage but also employment, business, education, social relations. It is also a typical market problem of optimizing supply and demand. Below we illustrate the problem by a simple example of optimizing the employment of four students in four universities. The procedure may be applied in many areas, including marriage, of course.

Let us have four post-doc candidates, call them Peter, Paul, Juan and Maria, aspiring to best universities, and assume there are four prominent universities, say those of Padova, Heidelberg, Oxford and Kiev, opening exactly the same number of post-doc positions/scholarships, one position in each. The students are not of equal capacity. The universities obviously want the best among the students, while the students do not care too much about the choice. 

\begin{figure}[H] 
		\centering
	\includegraphics[scale=1.3]{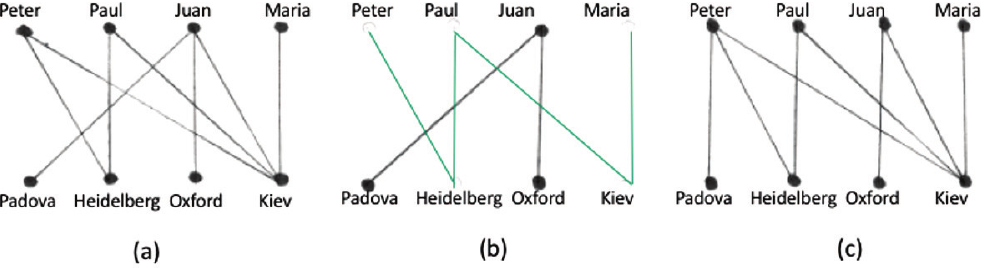}
	\caption{Visualising Hall's "marriage theorem" (demand vs. supply).}
	\label{Fig:Hall}
\end{figure}

In Fig. \ref{Fig:Hall} a biparticle graph, the students are on the top and the universities are at the bottom. A student and a university are connected if the university wants to have that student. For example, Kiev will invite any student, so it is connected with all four applicants, as shown in Fig. \ref{Fig:Hall}, left panel. 

Hall's theorem suggests the following optimal matching:
suppose $F$ is a biparticle graph ($A, B$). There is matching covering $A$ iff for every subset $X\subseteq A,\ \ N(X)\geq|X|$, where $N(X)$ is the number of neighbours of $X$. 

Let us illustrate Hall's condition in the following way: for a set of $n$ universities denote by $m$ the number of students that at least one of these universities want to have. If $m>n$ for every list of universities, than a matching is possible. Otherwise it is not. 


In Fig. \ref{Fig:Hall}, the middle panel highlights two universities (Padova and Oxford), but only one student, Juan is wanted by either one. Thus, since $1<2$, the matching fails. A solution: suppose Padova University's Council will invite Pietro instead. Then matching is successful and every student gets a position. 

Generally speaking, the number of students and positions does not need to be equal. If, for example, there are 10 students and 4 positions, and one wishes to fill every position, one can still use Hall's theorem, however in this case not every student will be granted a position.      

 As argued by Barab\'asi \cite{Barabasi}, this problem has much in common with matching optimally the supply and demand. For example, it works when there is a certain number of aspirants applying for a position at a company (university etc.) and that company (university etc.) has a finite number of vacancies to be filled by best aspirants, see also  Ref. \cite{Marriage}.

 In what follows we will be interested in the relations a couple of opposite sex. NB: In general, "sex" refers to the biological differences between males and females, such as the genitalia and genetic differences. "Gender" is more difficult to define, but it can refer to the role of a male or female in society, known as a gender role, or an individual's concept of themselves, or gender identity.

\section{Conjugal life}\label{Sec:Conjugal}
In this Section I investigate the creation and evolution of a family, a married couple -- nucleus of any society. 

Marriage is usually preceded by a period of "pairing", {\it i.e.} search for the optimal partner, see Sec. \ref{Sec:Hall}.  

I consider several scenarios illustrating conjugal life (many more are credible!). The dynamics $V(r,t)$ of the couples' life will be shown as function of two variables -- distance between the actors $r$ and time $t$, as well as their "worldline" (actually, a 2-dimensional curved string/band) embedded (living in) a 3-dimensional space-time world, see Sec. \ref{Sec:Wlines}.

In the following Subsection we apply the successful model of inter-particle interactions known in high-energy physics, based on the so-called Cornell potential \cite{Cornell}. In Subsection \ref{Models} we use several empirical functions to model the conjugal life. Many more options are possible, e.g. those attached to milestones in couples' life, taken from the treasury of world art and literature, see Appendix B (Marriage quotes, aphorisms).    
  
 \subsection{Toy model}\label{Models}
In physics, the interaction between two charged particles, {\it e.g.} quarks, gluons is well described by the so-called Cornell potential \cite{Cornell}, balancing between attraction at large distances $r$ and repulsion at short distances:
\begin{equation}
V(r)=ar-b/r,
\end{equation}
where $a$ and $b$ are parameters. I apply this model to a married couple, extending it by introducing time dependence in $a$ and $b$, different for the partners, $a\rightarrow a_m(t)+a_f(t)$ and $b\rightarrow b_m(t)+b_f(t)$. Also, I add a background term $c(t)$ accounting for any external influence. Thus, the potential becomes

\begin{equation} \label{Eq:Cornell}
V(r,t)=[a_m(t)+a_f(t)]r-[b_m(t)+b_f(t)]/r+c(t).
\end{equation}      

\begin{figure}[H] 
		\centering
	\includegraphics[scale=1.2]{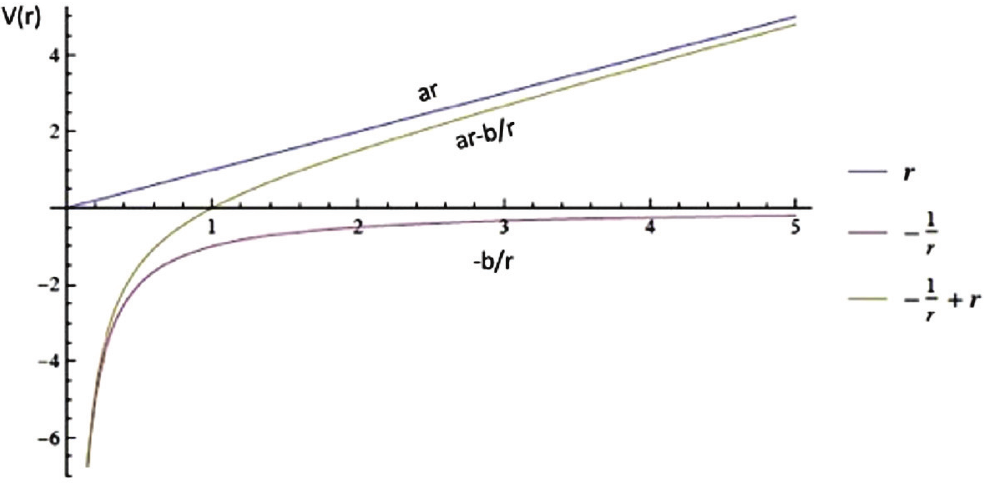}
	\caption{Conjugal relations following the "Cornell" potential Eq. (\ref{Eq:Cornell}) with time-independent (for the moment) parameters $a$ and $b$.}
\label{Fig:Cornell}
\end{figure}

\begin{figure}[H] 
		\centering
	\includegraphics[scale=0.95]{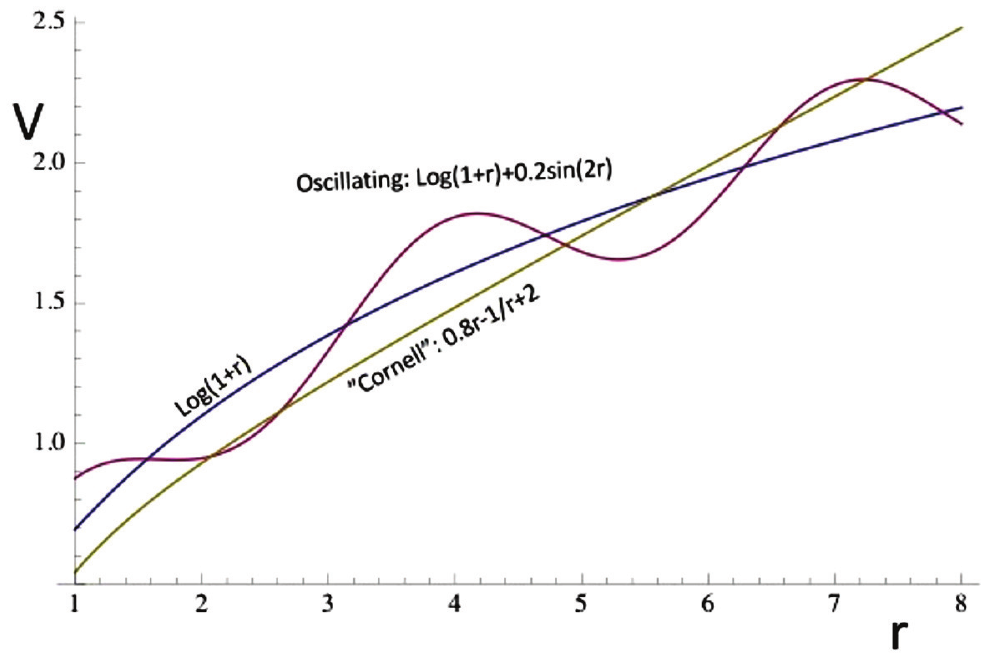}
	\caption{Simple (and optimistic) trends in conjugal relations; the "Cornell model" closely follows logarithmic rise. Such a monotonic behaviour may be perturbed by small oscillations as in the "Swedish family" of a 1973 TV miniseries written and directed by Ingmar Bergman, starring Liv Ullmann and Erland Josephson. 
	 Their matrimony is a sequence of attraction and repulsion caused by a mixture of common intellectual interests, sex, frustration etc., that may be described by a sinusoid as in Figs. \ref{Fig:Cornell1} and \ref{Fig:Cornell_Sin} (right), depending both on distance $r$ and time $t$.
Pessimistic scenarios, {\it i.e.} those with degrading or interrupted/ruined relations are not considered here.}	
\label{Fig:Cornell1}
\end{figure}

The interaction between individuals, similar to physics, is a function of the internal (inherent) properties of the individuals and their interrelation, both depending on time and relative distance.
 

To be specific, below I concentrate on the basic binary systems, that  formed by a male and female. My choice is motivated by: 1) the importance of the family as the basic cell in any society, 2)~similarity with the physical world: attraction between opposite (electric or magnetic) "charges" and repulsion between like charges. 

The above model is only an approximation to reality: manifestation of masculine or feminine characters, respectively by men and womenfolk is not as unique as for electric/magnetic charges. Even within traditional sexual relations, men may be endowed with feminine features and vice versa. By this I mean psychology, not physiology (biology). (I avoid the delicate and disgusting subject  of "erroneous" inborn gender and its correction ("repair") by surgical intervention, popular in certain media).
I mean something simple and obvious: opposite characters attract compensating something they are short of; usually we avoid/reject what we dislike in ourselves, or we miss. Passive males usually chose an active female and {\it v.v.}. This "compensation mechanism" in human relations was subject of numerous studies, see e.g. \cite{Otto}.

The variety of tempers of the individuals (encoded in $a$ and $b$) may be modelled by replacing their sum with 
\begin{equation}\label{Eq:lambda}
A=\lambda a+(1-\lambda)b, 
\end{equation}
where $\lambda$ is a measure of masculinity (or, symmetrically, feminity), varying between $0$ and $1$. This formula accounts for complementarity, important in balancing attraction and repulsion, assuming that $a$ and $b$ correspond to opposite (or at least different) tempers (see \cite{Otto}). As noted, a balanced couple is that where the partners compensate the shortage/excess of their inborn qualities (egoism-altruism, openness-insularity,  optimism-pessimism, practicality-dreaminess, etc.). The equilibrium may be regulated by the parameter $\lambda$ in Eq. \ref{Eq:lambda}.
Available literature (see Appendix, Sec. \ref{Sec:Quotes}), folklore and daily observations offers innumerable examples of binary relations as functions of time and space.

\begin{figure}[H] 
		\centering
	\includegraphics[scale=0.35]{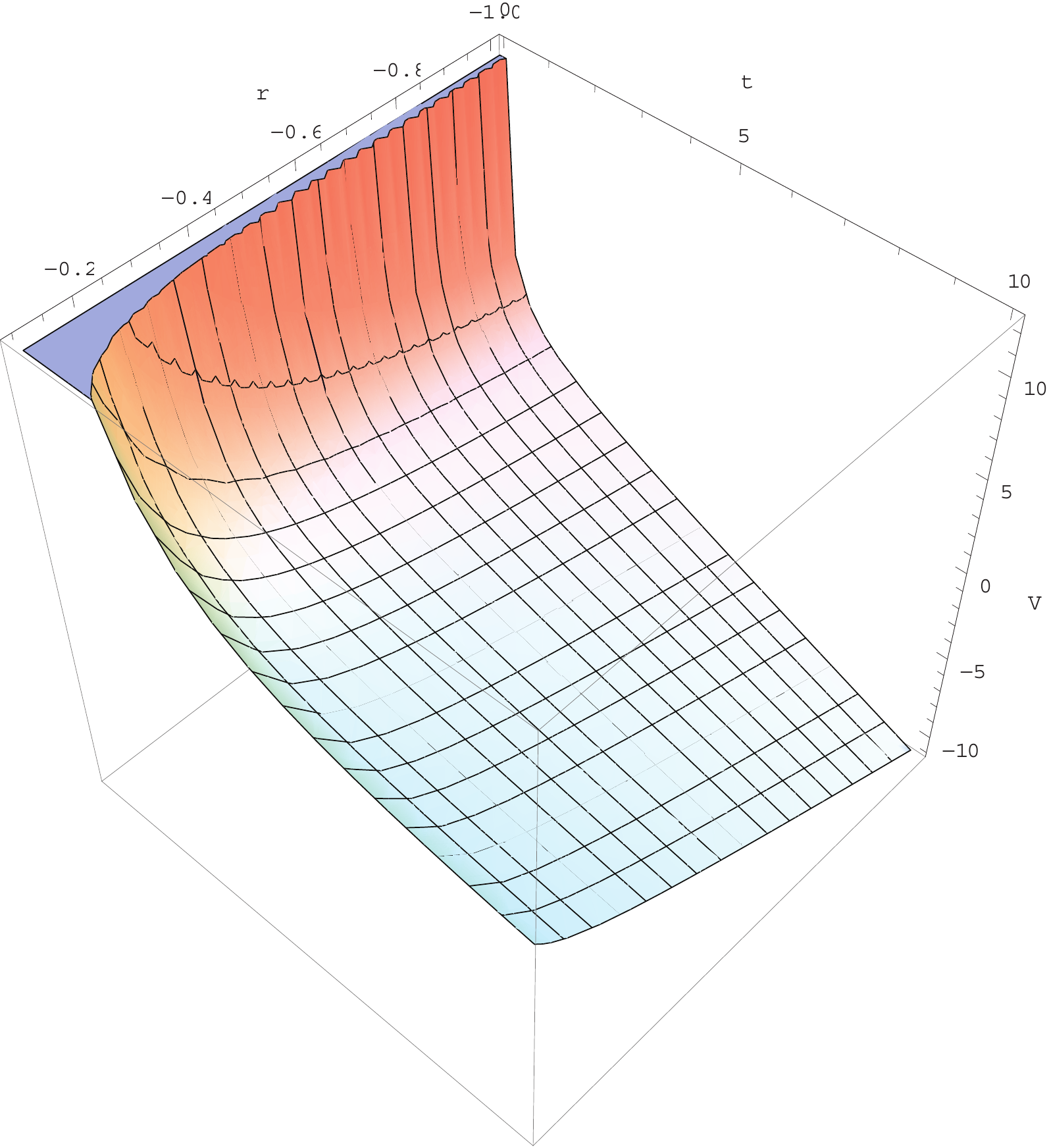}
	\includegraphics[scale=0.6]{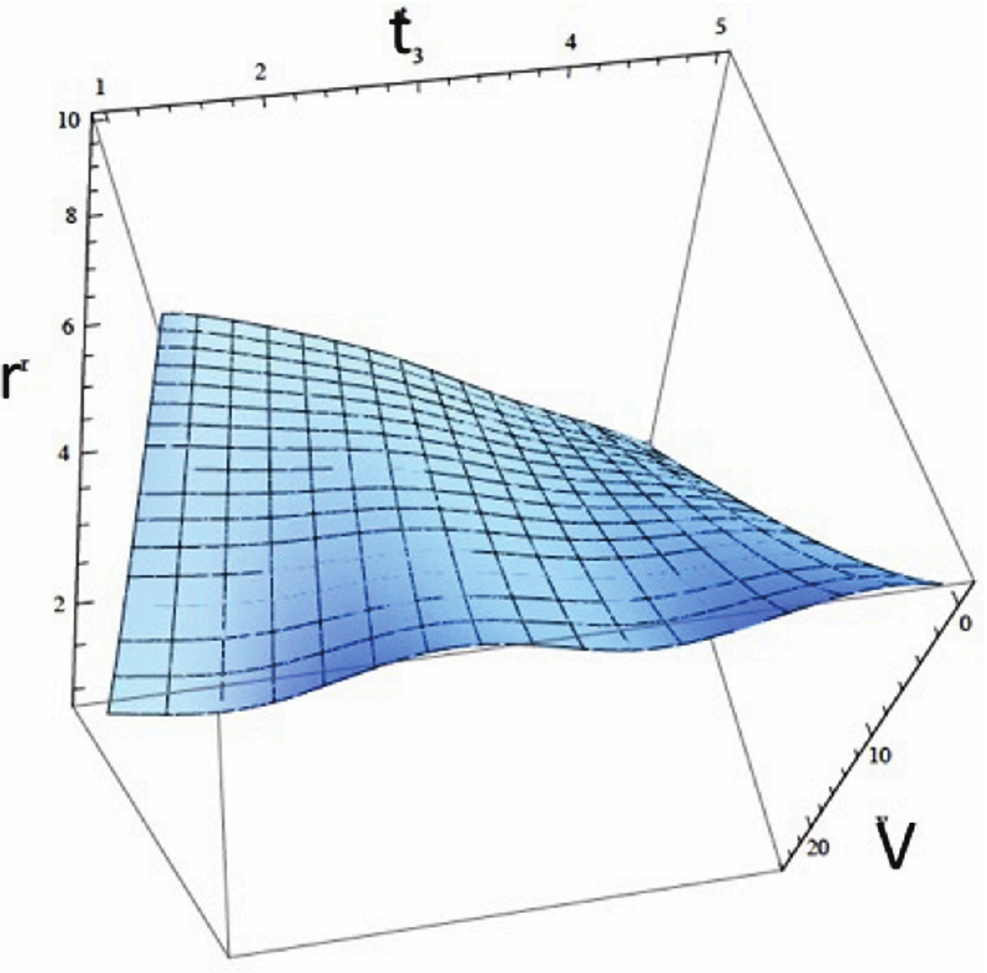}
	\caption{Two-dimensional plots of Figs. 7 and \ref{Fig:Cornell1} are generalized to 3D by introducing, apart from distance, also time dependence of the parameters $a\rightarrow a(t),\ b \rightarrow b(t)$ and $c\rightarrow c(t)$. NB: The right-hand icon shows also mild oscillations with time (cf. Fig. \ref{Fig:Cornell1}).}
	\label{Fig:Cornell_Sin}
	\end{figure}
	
\begin{figure}[H] 
		  \includegraphics[scale=0.7]{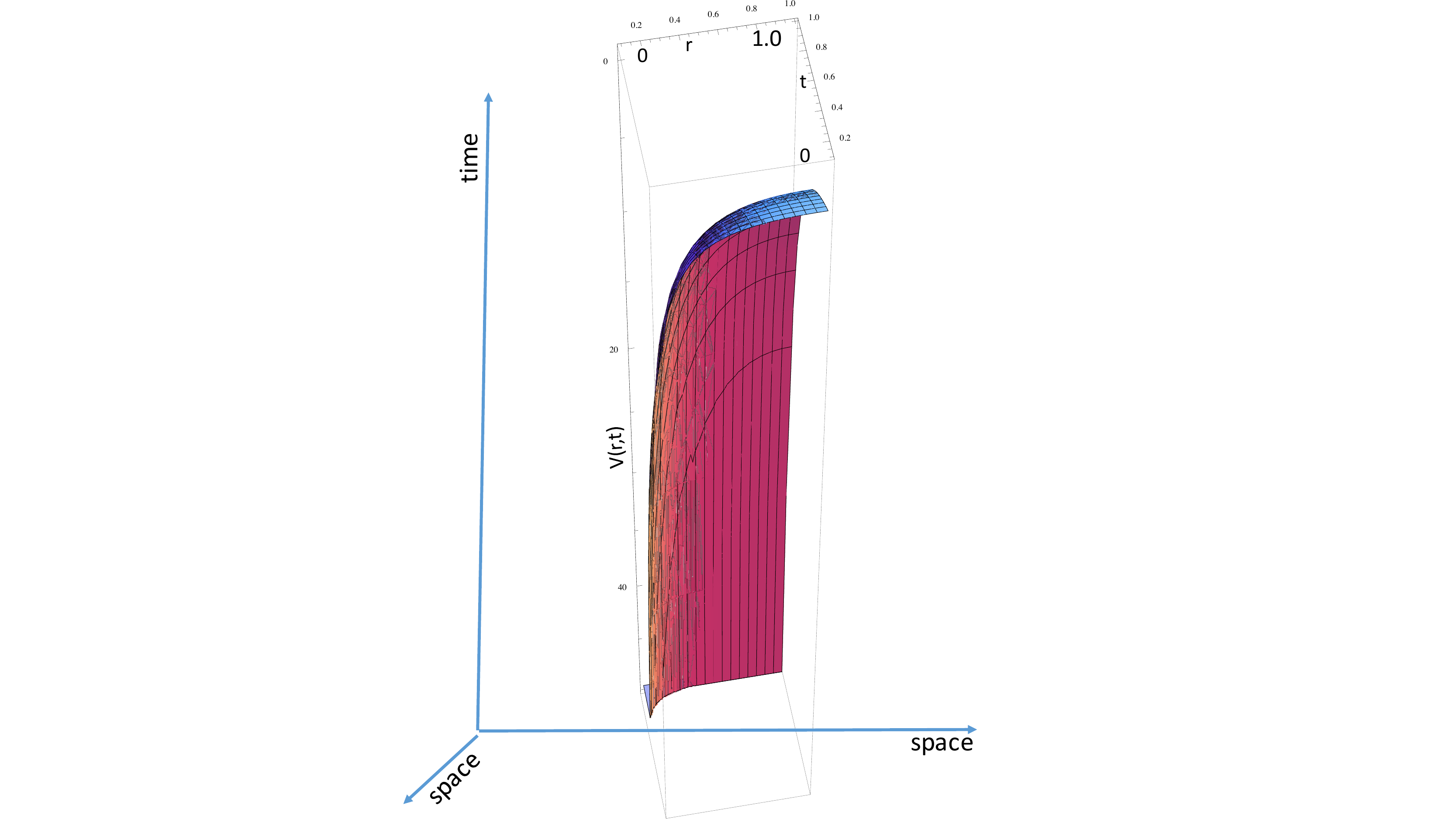}  
\caption{Rescaled surfaces (bands, strips) of Fig. 9 embedded in an external coordinate system $(x,y,t)$ live there string-like, the string’s elements (here, a married couple, labelled with $m$ and  $f$) interacting continuously, sweeping a world-sheet (brane) in the external $3$-dimensional space-time (cf. Figs. 2 and 3). This simplified case aims to illustrate the idea.}
	\label{Fig:Embed}
   	\end{figure}

 Figs. \ref{Fig:Cornell1} -- \ref{Fig:Embed} illustrate scenarios of a couple's life. Note an interesting phenomenon, known in the microworld, and related to repulsive forces at small distances $r$ known as {\it fall to the center} or {\it ultraviolate divergence}. The problem was extensively studied in relativistic quantum field theory, and cured by {\it renormalization technique} \cite{Bog}. A popular review can be found {\it e.g.} in 
 Ref. \cite{Kosyakov}. In a similar way, two persons (here a married couple), lean towards each other up to a certain distance corresponding to $V(r)=0$ in 
 Fig. \ref{Fig:Cornell1}. This is understandable: each human being has his own innermost "privacy territory", closed to external intervention. This limit is "case-dependent", but it necessarily exists! Violated, i.e. penetrated by an outsider (intruder), the core/sole may be damaged and the individual may loose his/her {\it ego}. This is similar to the materialistic microword: a particle looses its identity if an external agent penetrates the barrier of critical repulsion, see  Ref. \cite{Kosyakov}. 
 
 \subsection{Units and scales}\label{ssec:Units}

In moving from the material ("body") to the spiritual ("soul") world, we are facing the delicate problem of units, indispensable in natural sciences. Without going into details, let me only mention three options: classification of quantitative notions, comparative ones and quantitative, called also metric (contrary to the previous two, defined as "topological"). Let us try to be "metric" as far a possible, relying on the dominant logical, causal rather than chaotic behaviour of people. Causality implies that any effect is preceded (caused) by its cause. On the other hand, any physical action, including mechanical motion to large extent is induced by emotions, motivation. In other words, a {\it measurable} action, may be related/reduced, in some non-trivial way to- or derived from accompanying emotions.     

The toy model of Subsection \ref{Models} is an attempt to do so: translate ("materialize") feeling by using a familiar coordinate system.
The linear dependence on $r$ may be replaced by more complicated functions, but the above "Cornell" form is a convenient way to demonstrate the idea. By choosing $a,\ b,\ r,\ t$ as variables, we must define their dimensions as well as that of the "potential" $V(a,b,r,t)$. 
While the dimensions of $r$ (distance) and $t$ (time) are obvious, ({\it e.g.} meters and seconds), the intensity of "feelings", $a$ and $b$, are to be defined. A new unit may be introduced. In choosing the scale one may follow the definition of temperature, scaled to boiling and freezing points. For example, set $100$ degrees as the upper ("boiling") value of $V(r,t)$ (marriage) and $0$ as the "freezing" point, {\it e.g.} associated with separation (divorce etc). Actually, it is common to characterize feelings using "thermal" vocabulary, ranging from frozen relations ($0$ degrees), though cold, cool, worm, hot relations, arriving to boil (blow up) at $100$ degrees.

 In this section we have set down a framework admitting of various dynamical inputs, visualized by simple semi-quantitative examples. Common is the start: a marriage occurs when two individuals decide to connect their world lines. The subsequent evolution may follow different avenues. The predictive power of our formalism depends on the dynamics fed in. The main source of information comes from  observations (empiric), literature, inspiration, etc. Common and definite are the "initial conditions": people marry when they prefer to live together rather than separately -- attraction dominates over repulsion. 

 \section{Conclusions, perspectives and open questions}\label{Sec:Concl}
In this paper I attempted the almost impossible: to combine irrational/indeterministic human behaviour with rational/deterministic laws of physics.
The conclusion of this study are manifold. On the one hand, reproducing world lines of known people extended by relevant social network and genealogy, given the huge amount of "data" in the literature, is more than just amusing entertainment: it is instructive and useful not only for history but also as a check of the method presented in this paper, evolving towards useful applications and predictions. 

 Interesting is the next step involving multiply interacting networks fitted to more than two actors. In any case, understanding binary systems is indispensable to progress towards more complex collective systems including possible critical phenomena, with phase transitions in social systems. Van der Waals forces and the relevant equation, similarly to binary systems discussed in Sec. \ref{Sec:Conjugal} are based on attraction and repulsion between the constituents, offering many possibilities of their application in sociophysics. Simple semi-quantitative examples in that Section (Figs. \ref{Fig:Cornell}--\ref{Fig:Embed}) are meant merely to illustrate the basic ideas.

Promising are studies of an increasing number of worldlines/tubes, including continuum (merge of WLs). With the advent of computational and storage capacities, infinitely large manifolds of WLs, coming close and interacting multiply, evolving towards a continuous $3$-dimensional bulk of world history, parametrized numerically of by phenomenological models. Modern computing and storage capacities offer perspectives to handle the interaction of large numbers of encoded  world sheets.

In studying the behaviour of a large number of individuals one faces two kinds of hierarchic systems: totally hierarchic (vertical, as in a WL) and completely democratic (horizontal networks) systems The real world is a mixtures of two.

The above-mentioned bipolarity (Western "democracy" vs. Eastern "administrative command system") has a parallel with the title of the present paper: while {\it networks} correspond to democracy, {\it worldlines} are hierarchic. 


In perspectives, worldlines may play an important role in descriptive history. By this I mean a panoramic view of the evolution of the society, including WLs of individuals and groups, societies etc., as well as their interaction/intersection at various levels and forms. 
Realization of such a huge "bank of world lines" technically was incredible in the past, but now, with the advent huge computation and storage capacities, it may be realized! 

The fate of an ethnic/linguistic minorities in alien environment may be modelled by a drop of oil in water. Chances for its survival/assimilation (by solution) depends on its homogeneity and surface tension of the drop and aggression of the medium, the surrounding liquid. The theory of percolation may mimic contacts 
and flow across borders. We intend to study these phenomenon with applications to familiar examples of big diaspora: Armenian, Russian, Jewish, Chinese, Hungarian, Spanish etc.

\section*{Ackowledgements}
I thank A. Zhokhin for numerous  discussions and Yu. Shtanov for useful remarks.
\section*{Appenix A. Types and dimensions of homophile} \label{Sec:AppI}
\begin{itemize}
\item{\bf Baseline vs.inbreeding.}
One distinguishes between baseline homophile and inbreeding homophily. The former is the amount of homophily that would be expected by chance given an existing uneven distribution of people with varying characteristics, and the second is the amount of homophily over and above the expected value  \cite{Homophily}.

\item{\bf Status vs. value.}
Different are the status homophile and value homophile: individuals with similar social status characteristics are more likely to associate with each other than by chance. "Status" includes both ascribed characteristics like race, ethnicity, sex, and age. In contrast, value homophile involves association with others who think in similar ways.

\item{\bf Race and ethnicity.}
Social networks may be affected by race and ethnicity, which account for the greatest proportion of inbreeding homophile. Smaller groups have lower diversity simply due to the number of members, and this tends to give racial and ethnic minority groups a higher baseline homophile.

\item{\bf Sex/gender.}
As to sex and gender, baseline homophile of networks is relatively low compared to race and ethnicity. Men and women frequently live together, and are both large and equally-sized populations.

\item{\bf Age.}
Most age homophile is of the baseline type. For example, the larger age gap someone had, the smaller chances that they were confided by others with lower ages.

\item{\bf Religion.}
Homophile based on religion is due to both baseline homophile  and inbreeding.

\item{\bf Education, occupation and social class.}
Parents account for considerable baseline homophile with respect to education, occupation, and social class.
\end{itemize}

\section*{Appendix B. Marriage quotes, aphorisms} \label{Sec:Quotes}

Literature and folklore offer an inexhaustible source to guide empirical world lines, see Sec. \ref{Models}. Below is a short selection of aphorisms. Many more were collected and commented by Leo Tolstoy, see \cite{Leo}.    

{\bf Fran\c coise Sagan}:
 
All marriages are successful. Difficulties begin when living together begins.  

{\bf Leo Tolstoy, "War and piece"}:


Les marriages se font dans les cieux (Die Ehen werden im Himmel geschlossen); 

{\bf Leo Tolstoy, "Anna Karenina"}: 

 All happy families resemble one another, each unhappy family is unhappy in its own way. 
 

{\bf Maria Ebner von Eschenbach}:

Marriages are made in heaven, but they do not care that they are successful. 


{\bf A.I. Kuprin}:

     
Separation for love is the same as the wind for fire: it extinguishes a small love, and inflates a large one even more).

{\bf Friedrich Wilhelm Nitzsche}:

If the couple did not live together, successful marriages would occur more often.

{\bf Folklore}:  

 https://ru.citaty.net/tsitaty/452612-mariia-fon-ebner-eshenbakh-braki-sovershaiutsia-na-nebesakh-no-tam-ne-zabotiatsia/  


Out of sight, out of mind (Aus den Augen – aus dem Sinn).

Far from eye, far from heart.

It's one step from love to hatred.

\section*{Appendix C. Simple Wolfram Mathematica codes} \label{Wolfrem} 
For readers' convenience and easement, below we quote several simple Wolfram Mathematica codes used in Subsec. \ref{Models}, Figs. \ref{Fig:Cornell}--\ref{Fig:Embed}. The toy models are intended merely to illustrate the idea, 
opening the way to more sophisticated applications.   

 {\bf Cornell Conjugal}:
\begin{verbatim}   
a1 = a2 = 1; b1 = b2 = 1; al1 = a1/t; al2 = a2/t; bl1 = b1; t; bl2 = b2*t; cl = 0;
Plot3D[{-(al1 + al2)/r + (bl1 + bl2) r + cl}, {r, -0.1, -1} , {t, 0.1, 10}, AxesLabel -> {r, t, V}, BoxRatios -> {1, 1, 1}]
\end{verbatim}

{\bf World sheet; hierarchic coordinates}:
\begin{verbatim}
a1 = a2 = 1; b1 = b2 = 1; al1 = a1/t; al2 = a2/t; bl1 = b1; t; bl2 = b2*t; cl = 0; 
Plot3D[{-(al1 + al2)/r + (bl1 + bl2) r + cl}, {r, -0.1, -1} , {t,  0.1, 10}, AxesLabel -> {r, t, V}, BoxRatios -> {1, 1, 10}]
\end{verbatim} 
\newpage
{\bf Cornell3D}
\begin{verbatim}
 a1 = a2 = 1; b1 = b2 = 1; al1 = a1*t; al2 = a2; bl1 = b1; bl2 = 
 b2/t^(0.1); cl = 0;Plot3D[{-0.1 (al1 + al2) r + 10 (bl1 + bl2)/r + cl}, {r, 2, 5} , {t,5, 10}, AxesLabel -> {r, t, V}, BoxRatios -> {1, 1, 1}]
\end{verbatim}

\begin{verbatim}
 a1 = a2 = 1; b1 = b2 = 1; al1 = a1*t; al2 = a2*Sin[3 t]; bl1 = 
 b1*t; bl2 = b2/t; cl = 0;
  Plot3D[{-(al1 + 2 al2)/(r + 1) + 0.5 (bl1 + bl2) r + cl}, {r, 1, 10} , {t, 1, 5}, BoxRatios -> {1, 1, 1}, 
   AxesLabel -> {"r", "t", "v"}]
   
      c = 0; a = 1; b = 1; f1 = a*r; f2 = -b/r; f3 = 
 f1 + f2 + c; Plot[{f1, f2, f3}, {r, 0, 5}, 
 PlotLegends -> "Expressions"]
 \end{verbatim}


\end{document}